# Context-aware Adaptive Personalized Recommendation: A Meta-Hybrid

Peter Tibensky, Slovak University of Technology, Ilkovicova 2, Bratislava, Slovakia, peter.tibensky@icloud.com

Michal Kompan, Kempelen Institute of Intelligent Technologies, Mlynske nivy 5, Bratislava, Slovakia, michal.kompan@kinit.sk, ORCiD 0000-0002-4649-5120

**Abstract.** Recommenders take place on a wide scale of e-commerce systems, reducing the problem of information overload. The most common approach is to choose a recommender used by the system to make predictions. However, users vary from each other; thus, a one-fits-all approach seems to be sub-optimal. In this paper, we propose a meta-hybrid recommender that uses machine learning to predict an optimal algorithm. In this way, the best-performing recommender is used for each specific session and user. This selection depends on contextual and preferential information collected about the user. We use standard MovieLens and The Movie DB datasets for offline evaluation. We show that based on the proposed model, it is possible to predict which recommender will provide the most precise recommendations to a user. The theoretical performance of our meta-hybrid outperforms separate approaches by 20-50% in normalized Discounted Gain and Root Mean Square Error metrics. However, it is hard to obtain the optimal performance based on widely-used standard information stored about users.

## 1. Introduction

In the field of recommender systems, we recognize several techniques like collaborative filtering, content-based, knowledge-based, and demographic (Ricci, Rokach, Shapira, & Kantor, 2010). Furthermore, there are hybrid recommender systems combining approaches of the mentioned techniques to reduce the impact of their drawbacks and, on the other hand, strengthen their advantages to increase the performance of recommendation (Burke, 2002).

The typical approach is to select a recommender method that provides recommendations to every user in the same fashion. Nevertheless, users in the system behave differently, and the way they use the system varies. Historically, numerous recommender methods have been proposed – usually motivated by specific problems of application tasks. It is the "holy grail" of the field to have a recommender performing for all domains or user characteristics. Not surprisingly, no such method has been proposed, and various strategies perform better under various circumstances. In other words, each of the methods achieves different performance for different users depending on their interactions with the system. For instance, content-based methods do better for users rating similar items or cold-start users. Collaborative filtering increases the diversity of the recommendations. Therefore, using various methods whose choice depends on the user and contextual information may lead to improve recommendation performance.

This paper proposes a meta-hybrid recommender system that utilizes information about users and their context to select a method to generate recommendations. According to Burke (Burke, 2020) a meta-level recommender uses the output of one model as an input for another. The switching recommender switches between two or more recommenders based on a predefined rule. Our meta-hybrid combines these two approaches, which results in a trained classifier that utilizes a user context to select an optimal specific recommender method for a specific recommendation request and user. Various types of context are used based on the domain and the purpose of an application. We utilize the context of the user rating and items a user interacts with (e.g., date and time aspects of user's rating, diversity of genres). This information is extracted from the user's feedback and items he/she rated before. Next, a classifier is used to determine the most precise method for them. We also demonstrate the possibility of the use of contextual and preferential information. Thanks to this information, the classifier adaptively selects a method to provide the most precise recommendations.

In the evaluation phase, we explore three research questions, which aim at step-by-step exploring the proposed method performance. First, we explore the performance of the arbitrary recommenders approaches. Then the importance of the context model features is evaluated. Finally, we explore the performance of the meta-hybrid approach over the MovieLens dataset.

The main contributions presented in the paper:

- Proposing of novel meta-hybrid recommender which utilizes the user's context to select optimal recommender algorithm.
- Exploring the most important features of the user context model, which contribute to the prediction performance.
- The support of the proposed meta-hybrid idea by the theoretically optimal performance reachable by the proposed approach (which highly outperforms baselines).

## 2. Related works

Each of the recommender techniques suffers to some degree to various typical problems well-known in the recommenders' field. Often a hybrid recommender is used to reduce one or several of these.

The problem of the cold start can be present in a system in two forms. Firstly, if a new user starts using a recommender, no history can be utilized to provide relevant recommendations. Similarly, when a new item (e.g., news article) is added to the catalog, no interaction can be used for generating recommendations (Sammut, & Webb, 2017; Ricci, Rokach, Shapira & Kantor, 2010). Usually, the content-based approaches deal with cold-start better by definition as they require less user activity (Amatriain, & Basilico, 2016). These are used to search for similar items that have been in the system for a more extended period (Gaspar, Kompan, Koncal, & Bielikova, 2019).

The most popular combination of approaches used in the hybrid recommender is to utilize collaborative filtering and the content-based recommender (Walek, & Fojtik, 2020). The content-based approaches extract helpful information from the structured (Hernández-Rubio, Cantador, & Bellogín, 2019) and unstructured (Jain, Pamula, Ansari, Sharma, & Maddala, 2019) products' descriptions. Moreover, based on numerous research (Pradhan, Khandelwal, Chaturvedi, & Sharma, 2020; Agarwal, Sharma, & Katarya, 2019; Garg, & Sharma, 2016), the user reviews and the sentiment extracted from these texts is useful as the user-feedback information. When more structure metadata is available (e.g., movies), a preference for each distinct group can be modeled – as preferred genres, actors, etc. (Kaššák, Kompan, & Bieliková, 2016).

Depending on the domain, usually, the data used for the recommendations are very sparse. As there are often thousands of items in the catalog, it is clear that a user explores only a small part (Shalom,

Berkovsky, Ronen, Ziklik, & Amihood, 2015). This is usually true for the e-commerce domain. For instance, in the literature review by Danilova and Ponomarev (Danilova, & Ponomarev, 2016), 52% of hybrid systems were proposed in the e-commerce domain. As a result, often some additional information modeled in the user model is used to improve the recommendations (Berger, & Kompan, 2019).

The typical problem of several domains and the recommender algorithm also is overspecialization. This means that the recommender algorithm is highly focused on user preference and generates recommendations relevant to this and only this preference (Kotkov, Veijalainen, & Wang, 2018).

Another common phenomenon is called gray sheep, which mainly affects collaborative filtering approaches (Su, & Khoshgoftaar, 2009). Gray sheep are users whose preferences are often changing and do not correlate to any users group. As a reason, it is challenging to generate recommendations for them. Similarly, the black sheep (Zheng, Agnani, & Singh, 2017) users have rare preferences that differ from all users.

Several authors previously studied the problem of picking a specific algorithm in the domain of recommendation. Authors of (Aggarwal, 2016) proposed to use Multi-armed bandit (MAB) algorithms to deal with the challenge of new users and items constantly appearing in the system. The recommender system has to constantly adapt to these changes to provide relevant recommendations to each user. Felicio et al. (Felício, Paixão, Barcelos, & Preux, 2017) utilized MAB to select the right cluster for a user without having any feedback from him/her. First, they used matrix factorization to predict ratings of users who already have rated items and then cluster them using the Euclidean distance function. Then, they computed average ratings for every item and cluster. MAB then selects one of the clusters for an active user, and items with the highest rating are recommended. Active user's feedback is then used as a reward to recalculate MAB's prediction model. MAB is also used in many other recommender systems. For instance, the Amazon Stream service (Teo et al., 2016) utilizes Thompson sampling. Hariri et al. reported that MBA can adapt to changes in preferences, and multiple MAB algorithms can be used to maximize the average utility over each user's session (Hariri, Mobasher, & Burke, 2015).

Using a recommender method to recommend another recommender method is the next step. In (Cunha, Soares, & Carvalho, 2018), the authors utilized collaborative filtering to recommend a collaborative filtering method to provide the optimal precision of recommendations on an input dataset. The authors consider the input dataset to be active users and CF methods to be items to recommend. As far as the rating matrix is concerned, they model preferences using sub-sampling landmarks obtained from a random sample of instances, which determine the performance of algorithms on datasets.

A similar problem was researched by Ferrari (Ferrari, & Castro, 2012). Instead of recommendation, clustering algorithm selection for an input dataset was explored. This method uses meta-knowledge, which includes meta-data about a dataset such as *$log_2$* of the number of objects, *$log_2$* of the number of attributes, the proportion of binary, discrete and continuous attributes, and the correlation between continuous attributes. All of these meta-attributes do not require any semantics about the content of the dataset. After extracting the meta-attributes, the authors used the meta-algorithm to learn the relationship between the meta-attributes and the ranking. KNN provides the best result using the t-test metric.

The hybrid recommenders have been a standard solution for numerous problems of single methods as cold-start, overspecialization, sparsity, or any kind of "sheep" for a long time. In many domains, a combination of content-based and collaborative methods brings significant performance improvement. By proposing more and more recommender approaches, a combination of one-paradigm approaches seems to deliver improved results (e.g., using only collaborative approaches).

All of these approaches deal with the problem of strategy selection. While explore-exploit solutions like multi-armed bandit gradually improve their recommendation performance, recommending a recommender method and using meta-knowledge use some kind of information extracted from the dataset. The MBA methods gain a lot of researchers' attention in the last years. They usually require a lot of user activity and thus data. Moreover, the performance of such approaches is hardly approximated in offline settings.

The standard strategy selection methods are designed to select one method for the dataset or a user cluster. Many researchers and we believe that a step further is required to improve the performance. An optimal method is selected for each user. Moreover, a specific user under various circumstances should be served by different recommenders. For this purpose, a meta-hybrid seems to be a feasible approach. Similarly, the contextual MAB methods are an example of selecting a strategy based on user context (Edwards, & Leslie, 2020; Tekin, & Turgay, 2017).

## 3. Proposed method

The idea of the proposed method (Figure 1) is based on the usage of the meta-hybrid recommender. The meta-learner selects a specific recommender method for a user according to his/her actual context model. The user context model includes user feedback, information derived from feedback, and context. In this way, an optimal recommender list is generated and presented to the user.

We consider the selection of the recommender method to be a classification problem. In this task to a user is assigned a class representing the recommender method. The proposed meta-hybrid requires the user context model on the input to determine which recommender method provides the most precise results. The recommender algorithm consists of the following steps:

1. Construct user context model,
2. Predict recommender to be used (for each recommendation request, i.e., for the same user in various contexts, different recommenders may be predicted),
3. Recommend *n* items to the user using the predicted method.

### 3.1. User context model

The classifier utilizes information about users reflecting on their preferences and behavior. The user context model represents a vector containing these features. We propose to include three groups of attributes to be included in the user context model:

1. *Contextual attributes.* Contextual information contains meta-information extracted from the item descriptions, user's behavior, and background known to the recommender system.
2. *User preferences*. Preferential information is gathered from the user's explicit and implicit feedback and reflects the user's taste.
3. *User demography*. A user demography helps to group users with similar behavior and preferences.

The context models are regularly updated to capture the user behavior (and context) change.

As the method is evaluated in the domain of movies, the user context model includes domain-specific attributes, which can be easily replaced when applying to other domains. The complete list of the attributes used is presented in Table 1. Several attributes are closely related to the number of movies a user has rated. For these attributes, a normalized value is stored (Table 1, (n)).

It is clear that several attributes in the proposed user context model are multidimensional attributes. For instance, the histogram of user's ratings, or histogram of genres. These are included in the model as a whole (extending the dimensionality of the context model). In other words, if the user context model consists of *n* attributes, and there are 7.8k distinct keywords in the dataset, a vector of length 7.8k (each value represents a count of movies, rated by a user with a specific keyword) is added to the user context model as new attributes. This will result in the new length of the user context model - *n+7.8k.*

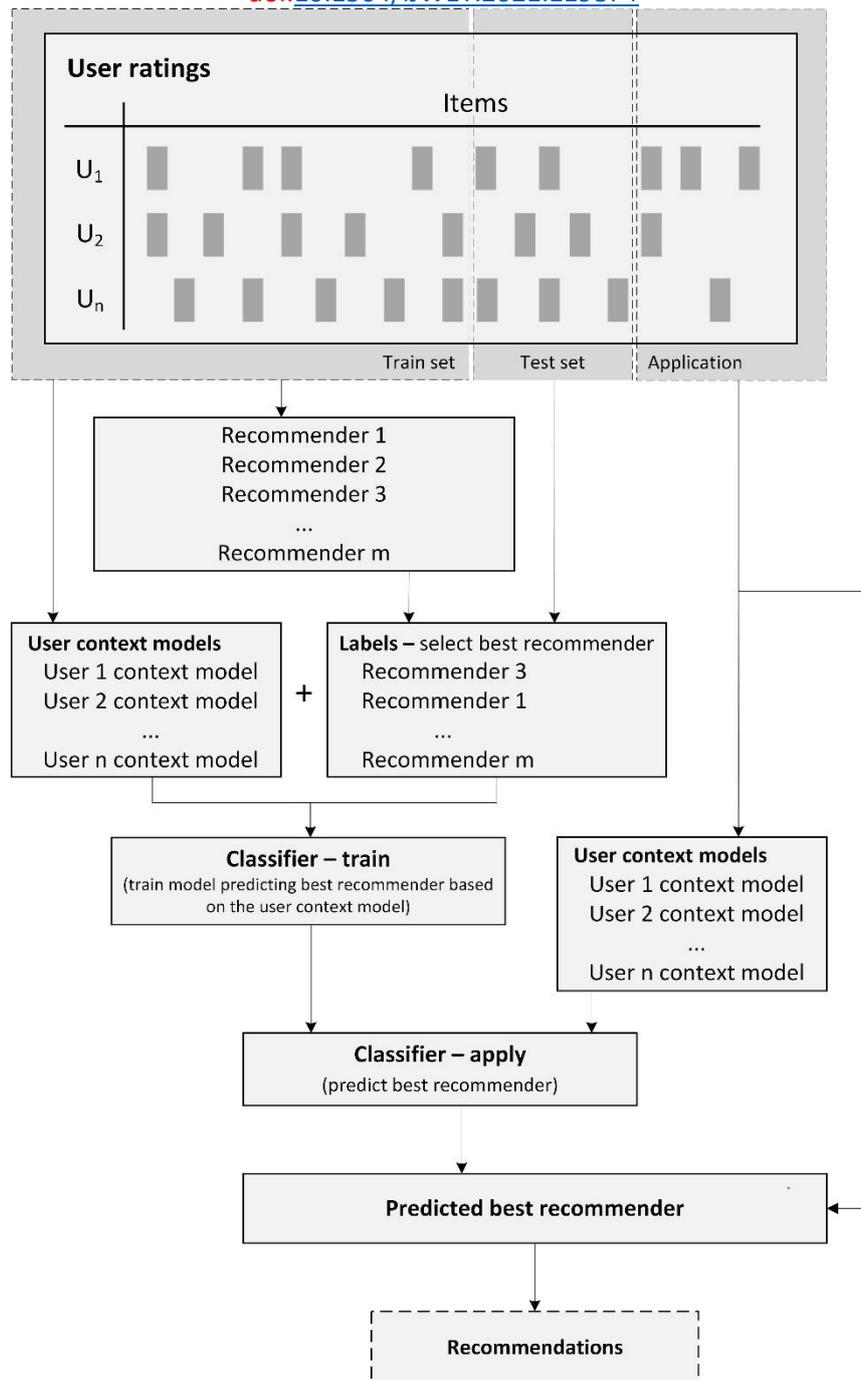

*Figure 1. The idea of the proposed approach. Firstly, the user context models are generated based on the historical data, and the best recommender (based on the nDCG (normalized Discounted Cumulative Gain) over Test set) for each user context model is selected (as a label). Next, the classifier is trained to be able to predict the optimal recommender algorithm based on the user context model. Finally, in the application phase, for a user to whom the recommendation is requested, his/her user context model is created, a prediction of the best recommender is performed, and the final recommendations are generated.*

### 3.2. Prediction step

The prediction step (Figure 1, Classifier - apply) represents an adaptive selection of a specific recommender approach based on a user's context model, which requires a recommendation (Algorithm 1). This prediction is based on the model - trained on historical data (Algorithm 2). We propose using the Random Forest classifier as it is well-known for its effectiveness on large datasets, satisfactory precision, and explainability. Each class represents one recommender approach. The set of recommenders used depends on the specific application and should be identified case-by-case; however, a diverse recommender should be used to utilize the potential of the proposed approach entirely.

In order to train a classifier for the optimal recommender prediction task, a train set have to be created. It is essential to point out that this is a step of the proposed method (not its evaluation). As the classification is a supervised learning approach, the labels have to be generated for each training instance. Subsequently, we train the classifier using a train set consisting of user context models labeled with classes (Figure 1, Classifier - train). The labeling is assigned by calculating the metric for each recommender method over each user context model and selecting one with the best nDCG (normalized Discounted Cumulative Gain). The labels are assigned as follows:

1. For each user represented by the user, the context model generates recommendations by all recommenders (Algorithm 2, lines 2-3).

2. Pick the best recommender for a specific context model based on the nDCG and assign it as the label (Algorithm 2, line 4).

*Table 1. List of attributes included in the user context model. The (n) indicates if an attribute is normalized by the number of movies a user has rated.*

| Group | Attribute |
|---|---|
| Contextual attributes | no. of items rated by a user |
| | histogram of user ratings (n) |
| | a variance of items metadata for items rated by a user |
| | a preferred hour and day of the week of user ratings |
| | no. of unique item categories, rated by a user |
| Preference attributes | histogram of genres of items, rated by a user (n) |
| | histogram of keywords for items rated by a user (n) |
| | length of movies, rated by a user (n) |
| Demography attributes | gender of a user |
| | user job |
| | user location |

Algorithm 1. Prediction step

```
Input: trained classifier model, userID
1. user context model ← get actual user context model (userID)
2. predicted recommender ← generate prediction (trained classifier model, user
                                                context model)
3. return predicted recommender
```

Algorithm 2. Classifier training

```
Input: set of candidate recommenders, train user context models
1. foreach instance in train user context models do
2.      foreach recommender in set of candidate recommenders do
3.            value_recommender ← nDCG(generate recommendations(recommender, instance))
```

```
4. train data ← (best recommender name(value), instance)
5. trainted classifier model ← train (train data)
6. return trained classifier model
```

As the recommender candidates, we consider several collaborative filtering, hybrid, and content-based methods. All of these methods are comparable to each other via nDCG, which we used to determine their performance. We selected these methods because they use different approaches to predict ratings. We assume their different properties will be reflected later in classification.

### 3.3. Recommendation step

The last step is to use a predicted recommender to generate recommendations for the target user (represented by his/her context model). The predicted recommender then generates recommendations based on the standard rating data or items' metadata. All recommender algorithms candidates (which we are using in the experiments) are based on the rating prediction. As a result, the final recommendation is generated as the Top-N items.

## 4. Evaluation
### 4.1. Data

We used the MovieLens 1M dataset containing over 1 million ratings[1] from approx. 6k users on 3.7k movies. For each movie, a list of metadata is available: title, year, genres. For each user, his/her age, gender, and occupation are provided. Since each user has at least 20 ratings (Figure 2), we randomly reduced the number of ratings for each user. Thanks to this "normalization," we obtained the so-called cold-start problem. As the user context model requires additional movie metadata, we enriched the MovieLens dataset with movie meta-data from The Movie DB[2]. As a result, for each movie, we stored: title, genres, plot, year, language, cast, keywords, average rating, budget, and profit.

---

[1] https://grouplens.org/datasets/movielens/1m/
[2] https://themoviedb.org

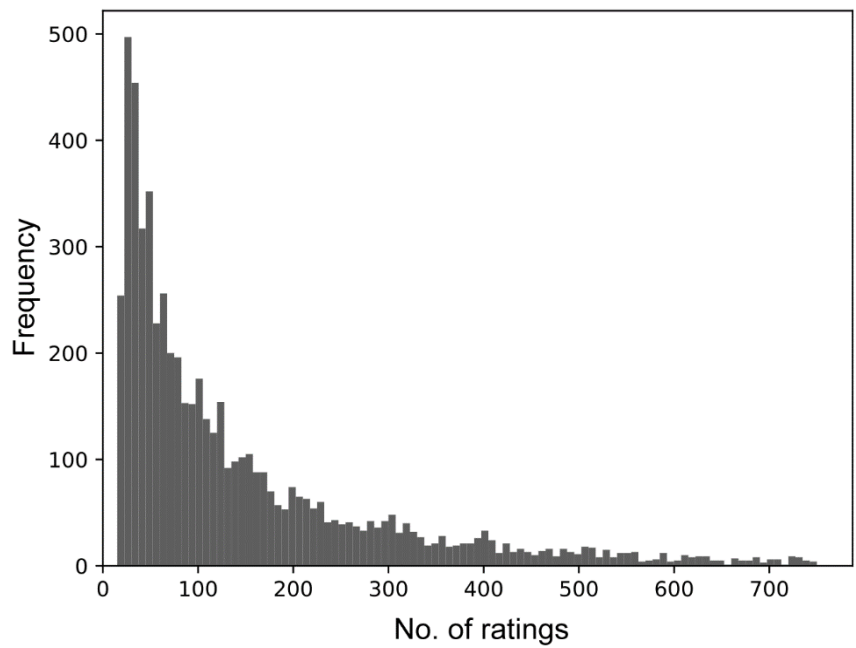

*Figure 2. Histogram of the user ratings in the MovieLens 1M dataset.*

The user context model (Table 1) consists of several attributes, which are one-hot encoded. In other words, a list of keywords, genres, etc., may significantly increase the number of attributes for each user context model (tens of genres and hundreds of keywords). In fact, this results in a worse-performing classifier. Because of this, we transformed these attributes – namely the histogram of keywords, and the histogram of genres, by the Principal Component Analysis as the standard approach for dimensionality reduction. By the rule of retaining most variance in the data in total 10 components for genres (Figure 3) and 15 components for the keywords were used.

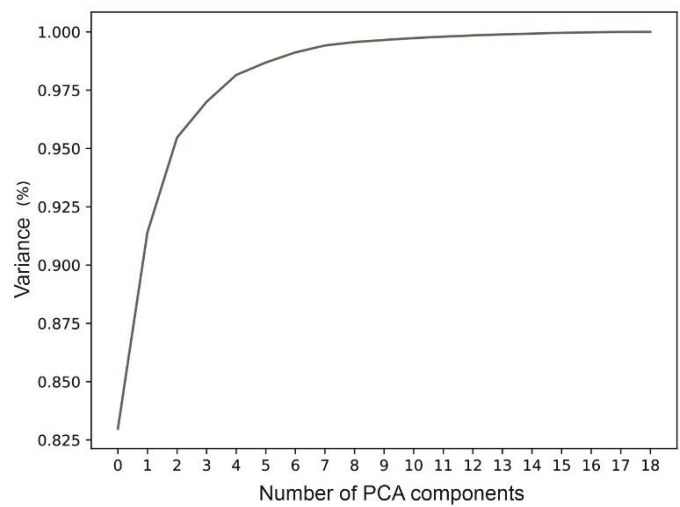

*Figure 3. The amount of variance explained by the specified number of PCA components for the genre.*

The final experiments were also conducted over the Yelp dataset[3] sample that contains user reviews on various businesses. Here, we have picked the reviews from the city of Austin and filtered out users with less than 20 ratings (reflecting the MovieLens dataset), ending up with 4.9k users, 17.8 items, and 250k ratings. For the content-based recommender, categories and attributes of the business were converted to binary vectors (there were no textual descriptions available).

---

[3] https://www.yelp.com/dataset

## 4.2. Methodology

To evaluate the proposed approach, we follow the standard offline evaluation methodology. However, as pointed in Figure 4, it is essential to notice the difference between the train and test set included in the method design and the train and test set from the evaluation point of view. The methodology was as follows:

1. Split the dataset into folds Train set - Hybrid approach $TR_h$ 70% and Test set - Hybrid approach $TE_h$ 30%
2. Split both folds into the train $TR_h - TR_r$, $TE_h - TR_r$ and test tests $TR_h - TE_r$, $TE_h - TE_r$
3. Use fold $TR_h - TR_r$ to train recommender methods
4. Determine which method for which users achieves the best nDCG score based on $TR_h - TE_r$
5. Train the classifier
6. Use $TE_h - TR_r$ to create a user context model and perform prediction for each user context model
7. Train predicted recommender methods using $TE_h - TR_r$
8. Evaluate generated recommendations to the ground truth based on $TE_h - TE_r$ set

Training the classifier and its testing is performed on folds containing different users to exclude any dependencies between the train and test phases.

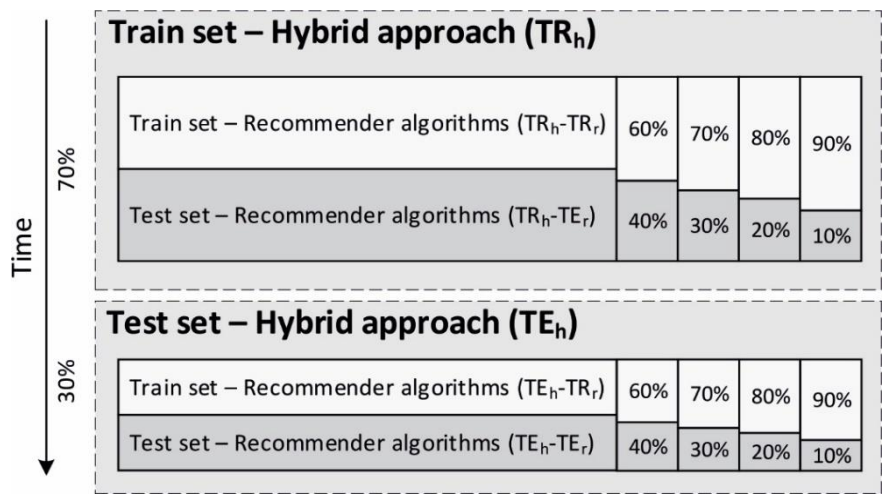

Figure 4. The idea of the train/test split for the evaluation. The available dataset is split 70:30 to the train ($TR_h$) and test ($TE_h$) set. As presented on Figure 1 also train and test sets are required for the proposed method itself. As a result $TR_h$ and $TE_h$ are further split to the Recommender algorithm train and test sets $TR_h - TR_r$, $TR_h - TE_r$ and $TE_h - TR_r$, $TE_h - TE_r$). In order to simulate recommendations for each user in several timestamps, various ratios are used: 60:40, 70:30, 80:20, and 90:10.

As the recommender algorithms candidates, we used several state-of-the-art collaborative recommenders from the Surprise[4] library: Baseline Only (Koren, 2010), Co-Clustering (George, &

---
[4] http://www.surpriselib.com

Merugu, 2005), Slope One (Lemire, & Maclachlan, 2007), SVD (Salakhutdinov, & Mnih, 2007), KNN Basic.

Moreover, we also use one hybrid algorithm LightFM[5] (Kula, 2015), and one content-based approach. The complete list of parameters used for each recommender algorithm is presented in Table 2. The values were obtained based on the randomized search implemented in the RandomizedSearchCV[6] method of the scikt-learn library. For the details on specific recommenders, please visit the Surprise[3] and LightFM[4] libraries.

*Table 2. A list of parameters used for each recommender algorithm.*

| Algorithm | Parameter | Value |
|---|---|---|
| Baseline Only | optimization | SGD |
| | user-based | True |
| Co-Clustering | no. of user clusters | 7 |
| | no. of item clusters | 5 |
| | no. of epochs | 30 |
| SVD | no. of factors | 20 |
| | no. of epochs | 30 |
| KNN | no. of neighbors | 50 |
| | similarity measure | Cosine |
| | user-based | True |
| LightFM | no. of components | 30 |
| | loss function | WARP |

We used two predictors for the optimal recommender algorithm selection - the Random Forest classifier and the Extreme Gradient Boosting (XGB). The optimal parameters for both classifiers are presented in Table 3.

*Table 3. A list of parameters used for predictors.*

| Algorithm | Parameter | Value |
|---|---|---|
| Random Forest | Bootstrap | True |
| | Criterion | Gini |
| | Max depth | None |
| | Min samples split | 3 |
| | Min samples leaf | 2 |
| | Max features | $\sqrt{no.\,of\,features}$ |
| | No. of estimators | 500 |
| XGB | Max depth | 150 |
| | No. of estimators | 500 |
| | Learning rate | 0.005 |
| | Reg. alpha | 0.05 |

---

[5] http://lyst.github.io/lightfm/docs/home.html
[6] https://scikit-learn.org/stable/modules/generated/sklearn.model_selection.RandomizedSearchCV.html

In order to evaluate the performance, we report standard, widely used metrics - Root Square Mean Error (Eq. 1) (Wang, & Lu, 2018), normalized Discounted Cumulative Gain (Eq. 2) (Clarke et al., 2008), and the Precision (Eq. 3) and Recall (Eq. 4) (Raghavan, Bollmann, & Jung, 1989).

$$RMSE = \sqrt{\frac{1}{|T|}\sum_{(u,i)\in T}(\hat{r}_{u,i} - r_{u,i})^2} \quad (1)$$

$$nDCG_p = \frac{\sum_{i=1}^{p}\frac{2^{rel_i}-1}{log_2(i+1)}}{\sum_{i=1}^{|REL_p|}\frac{2^{rel_i}-1}{log_2(i+1)}} \quad (2)$$

$$Precision = \frac{|Relevant \cap Recommended|}{|Recommended|} \quad (3)$$

$$Recall = \frac{|Relevant \cap Recommended|}{|Relevant|} \quad (4)$$

### 4.3. Results

*RQ1: What is the performance of recommender algorithm candidates on the MovieLens dataset?*

Our first research question aimed to determine how the distribution of the most accurate collaborative filtering methods per user on the MovieLens 1M dataset looks like. If the methods would perform similarly, it is useless to build a meta-hybrid based on these. The straightforward criterion is to compare algorithms based on the various user activity (i.e., the number of ratings in the train set).

As we can see (Figure 5), Co-Clustering leads over the other methods (when no cold-start is present). However, other methods achieve lower RMSE for 58% of users. RMSE difference between the best method and the second one is an average of *0.095*. The difference between the most accurate method and the others on an average of *0.175*. The SVD performs the best for the cold-start users. The average RMSE may be misleading; thus, we explored the histogram of the winning recommender algorithms. In other words, for each recommender candidate, we counted the number of users for which a specific approach was optimal. As we can see, the LightFM algorithm outperformed the other candidates. However, other algorithms were in sum better for 65.7% of users.

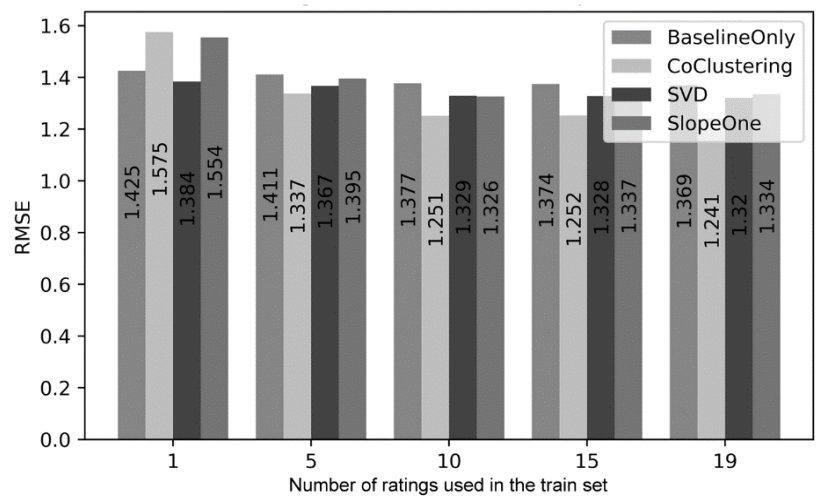

*Figure 5. Performance (Root Mean Square Error - lower better) of recommender algorithm candidates based on various train set sizes (i.e., number of ratings considered for a user).*

Next, we have focused on the recommender list quality. We have also included the LightFM and the content-based recommenders. Comparing with performances published on MyMediaLite[7] and Surprise, the difference is significant, and it is worth exploring adaptive method selection (Figure 6). As we can see, LightFM is the most successful recommender for the majority of users. On the contrary, the rest of the candidates are the best performers for significantly large user groups.

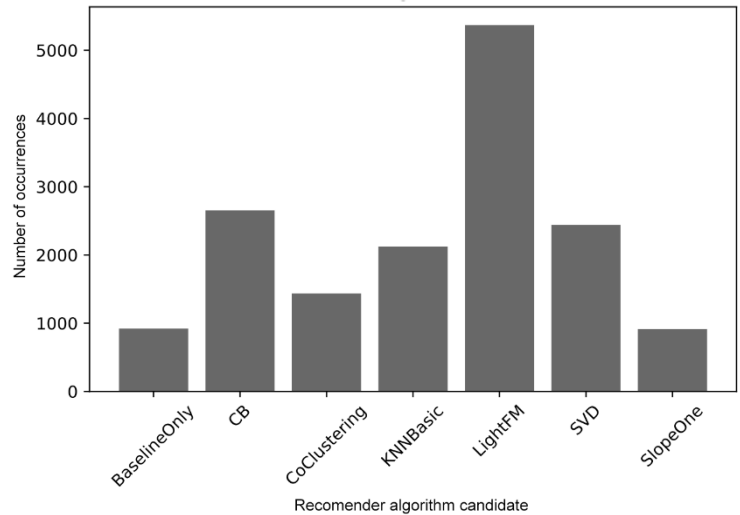

*Figure 6. The number of occurrences when each recommender algorithm candidate performs the best (based on the nDCG metric) for a user.*

RQ2: Which features included in a user context model are most important for the predictor?

Based on domain knowledge, we first considered the number of ratings and the distribution of ratings in classification. This is supported by numerous methods proposed for low activity domains in the literature. By analyzing the relation between the number of ratings and recommender methods' performances (RMSE), we identified the correlation that proves the existence of dependency between the number of ratings a user made and the method that recommends to the user with the lowest error.

Table 4 shows Co-Clustering achieves the lowest RMSE compared to others when a user has an average of 156 ratings. On the other hand, Baseline Only and Slope One perform their best when a user has 40 - 50 ratings.

*Table 4. Number of ratings when methods achieve the lowest RMSE.*

| Algorithm | Number of ratings | | | |
| --- | --- | --- | --- | --- |
| | Average | Quantile | | |
| | | 25% | 50% | 75% |
| Baseline Only | 40.66 | 11 | 25 | 52 |
| Co-Clustering | 156.05 | 26 | 72 | 188 |
| Slope One | 50.74 | 10 | 25 | 64 |
| SVD | 83.25 | 12 | 37 | 105 |

---

[7] http://www.mymedialite.net/

We also explored the importance of user context model attributes using the Random Forest classifier (Figure 7). As expected, based on the previous analysis - the occupation (i.e., the demography of the user) seems to be useless for the classifier (in the movie domain). On the contrary, the movie genres seem to be auspicious attributes set. We also explored the histogram of keywords and genre importances. As a result, it seems that used keywords are too specific and do not improve the prediction precision.

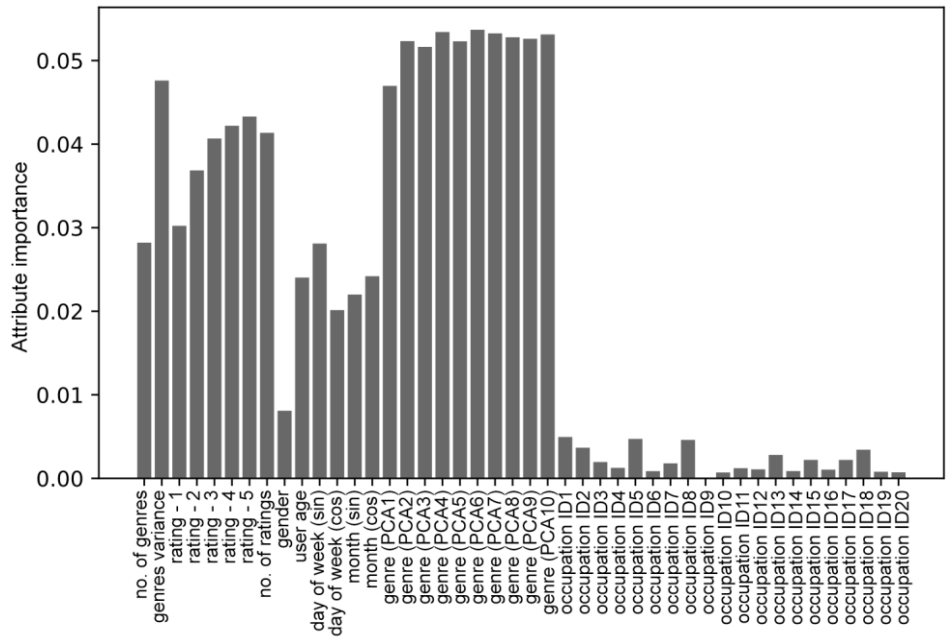

*Figure 7. The importance of selected user context model attributes based on the Decision Tree classifier.*

Obtained results indicate that for different domains, a set of distinctive attributes have to be explored. In some domains, attributes extracted from items' content may be necessary (e.g., keywords, sentiment); in other domains, user's activity-based attributes (e.g., number of ratings, purchases) or user demography may result in optimal results. Moreover, these attributes are dependent on the set of the recommender and its performance. In other words, diverse recommenders have to be used.

*RQ3: What is the performance of the proposed meta-hybrid approach in comparison to the recommender algorithm candidates?*

As the number of candidate recommenders may vary (hand by hand with their performance), we explored various settings and recommender candidates' groups.

Firstly, we focused on collaborative filtering candidates. Based on the literature overview and each candidate characteristic, we used: Baseline Only, Co-Clustering, Slope One, and SVD algorithms. These were included in the proposed meta-hybrid, and according to the methodology described in Section 4.2, the performance was evaluated.

The results show that the proposed meta-hybrid based on the user context model does not outperform the Slope One algorithm (Table 5). This is caused by the borderline performance of the classifier (based on the user context model), which is supported by the result of the Optimal hybrid (theoretically best solution) - which outperforms the second-best approach by 5.3% in the precision metric.

To understand the results, we further explored the meta-hybrid classifier and examined the wrong classified instances. We found out that error was produced mainly the wrong classification of Baseline Only class, and most of these miss classifications have resulted in the SVD class.

Secondly, we focused on the comparison of content and collaborative algorithm candidates. The idea of this experiment is based on standard widely used hybrid recommenders, which usually combine content and collaborative approaches.

*Table 5. The comparison of single collaborative recommender algorithms and the meta-hybrid which includes these algorithms. The Opt. hybrid refers to the optimal meta-hybrid, which selects the best algorithm for each recommendation.*

| Algorithm | Precision | | | Recall | | | nDCG |
|---|---|---|---|---|---|---|---|
| | P@3 | P@5 | P@10 | R@3 | R@5 | R@10 | |
| Baseline Only | .0861 | .0868 | .0866 | .0065 | .0111 | .0236 | .2078 |
| Co-Clustering | .0933 | .0910 | .0864 | .0081 | .0127 | .0229 | .2155 |
| Slope One | **.1036** | **.0998** | **.0933** | **.0088** | **.0138** | **.0255** | **.2373** |
| SVD | .0868 | .0839 | .0836 | .0067 | .0110 | .0211 | .1964 |
| Hybrid | .0993 | .0961 | .0901 | .0081 | .0129 | .0235 | .2240 |
| Opt. hybrid | **.1566** | **.1357** | **.1119** | **.0151** | **.0212** | **.0345** | **.3312** |

As we can see (Table 6) proposed meta-hybrid outperforms both KNN Basics and the Content-based approaches. However, it does not beat the pure LightFM approach - which is the hybrid recommender also. Similarly to the previous experiment, the theoretically optimal meta-hybrid scores the best. This indicates that the improvement of the classifier model and also the user context model would improve the performance.

Further analysis revealed that the most misclassified class was the LightFM (almost 1/3 of all instances). We believe that including more attributes in the user-context model would further improve the classifier. This is the main bottleneck, as the optimal hybrid (i.e., classifier with 0 misclassifications) will outperform all methods used as the partial recommenders.

The availability of additional contextual information varies across domains. For some domains, user-specific characteristics as mood are available (e.g., in the music domain). Often the weather, location, or other physical context are available, which may be relevant for some domains.

*Table 6. The comparison of content, collaborative, and hybrid recommender algorithms and the meta-hybrid which includes these algorithms over the MovieLens dataset. The Opt. hybrid refers to the optimal meta-hybrid, which selects the best algorithm for each recommendation.*

| Algorithm | Precision | | | Recall | | | nDCG |
|---|---|---|---|---|---|---|---|
| | P@3 | P@5 | P@10 | R@3 | R@5 | R@10 | |
| Content-based | .0703 | .0744 | .0751 | .0094 | .0156 | .0285 | .2210 |
| KNN-Basic | .1012 | .0989 | .0967 | .0082 | .0132 | .0265 | .2405 |
| LightFM | **.1579** | **.1506** | **.1407** | **.0200** | **.0318** | **.0582** | **.3594** |
| Hybrid | .1216 | .1187 | .1131 | .0133 | .0216 | .0396 | .2882 |
| Opt. hybrid | **.2343** | **.2112** | **.1785** | **.0299** | **.0448** | **.0744** | **.4933** |

To better interpret the results and observed patterns, we have used the Yelp dataset and the city of Austin. As we can see (Table 7), the results follow the trend of MovieLens dataset. All compared algorithms have scored very low, which can be explained by the dataset characteristic. Comparing to the MovieLens, the Yelp dataset is extremely sparse, where only 250k ratings are available on 17k items from 5k users. As we can see, the content-based approach has outperformed the collaborative methods followed by the proposed hybrid. The optimal hybrid again indicated that the idea of selecting the optimal algorithm based on the additional information could significantly improve the recommendations. However, we were not able to train the classifier sufficiently.

Table 7. The comparison of content, collaborative, and hybrid recommender algorithms and the meta-hybrid which includes these algorithms over the Yelp dataset. The Opt. hybrid refers to the optimal meta-hybrid, which selects the best algorithm for each recommendation.

| Algorithm | Precision | | | Recall | | | nDCG |
|---|---|---|---|---|---|---|---|
| | P@3 | P@5 | P@10 | R@3 | R@5 | R@10 | |
| Content-based | **.0127** | **.0113** | **.0104** | **.0032** | **.0046** | **.0083** | **.0460** |
| KNN-Basic | .0004 | .0008 | .0011 | .0001 | .0003 | .0007 | .0037 |
| LightFM | .0047 | .0034 | .0033 | .0010 | .0012 | .0022 | .0158 |
| Hybrid | .0065 | .0057 | .0061 | .0012 | .0018 | .0038 | .0255 |
| Opt. hybrid | .0154 | .0141 | .0144 | .0039 | .0039 | .0110 | .0616 |

From the computational requirements point of view is the proposed approach more consuming. Compared to the standard hybrid approaches, it introduces the classifier step on top of standard hybrid recommender requirements. The training phase of the proposed hybrid (without the optimizations) lasted 60x longer than training other single approaches. On the contrary, the theoretical improvement seems to be promising and worth despite increasing requirements. Moreover, when compared to not only single methods but to hybrid approaches, the proposed approach will differ only in the classification step (required only training phase).

## 5. Conclusion and future work

In this paper, we proposed a meta-hybrid recommender that uses contextual and preferential information about a user. The idea of the proposed meta-hybrid is to utilize actual user context and predict the recommender method, which will provide the most precise recommendations.

We have studied various settings and aspects of the proposed approach, which resulted in the following takeaways:

- For each recommender algorithm that was explored, an optimal algorithm exists for a group of users (sharing similar context).
- Various user context characteristics are essential for the identification of the best-performing recommender. In the movie domain, the number of ratings (i.e., the dataset sparsity), genre variance, user age were most important. On the contrary, the time-related information seems to be not necessary.
- The theoretical performance of the proposed meta-hybrid approach outperforms the single approaches by 20-50%. This indicates the considerable potential of our idea. However, based on the information available in the dataset used, we are not able to sufficiently train the classifier. This resulted in sub-optimal performance. To sum it up, further research is necessary to either use a more advanced classifier or extend the user context model with additional features.

We showed that we could identify a specific recommender algorithm by using an adaptive selection according to the user context model. According to our experiments, besides the user's feedback, context and content information are also important when selecting an optimal method. The availability of the relevant contextual information is crucial for predicting the optimal recommender method by the proposed idea and thus to provide an expected improvement over the specific recommenders recommender used as candidates.

Our experiments indicated that the classifier performance did not achieve a sufficient level, which resulted in a suboptimal result. However, the optimal meta-hybrid would outperform all partial recommenders used. In future work, we will aim to explore the meta-hybrid performance over various datasets. In fact, most of the standard used recommender datasets do not include context information. One of the most important directions is extending the user context model by features that will help to reduce the impact of outlier users and result in improved classification performance.

## Acknowledgments

This work was partially supported by the COST Action 19130. The authors wish to thank Peter Gaspar for his contribution to the experiments.

## References

Ricci, F., Rokach, L., Shapira, B., & Kantor, P.B. (2010). Recommender Systems Handbook (1st. ed.). Springer-Verlag, Berlin, Heidelberg.

Burke, R. (2002). Hybrid Recommender Systems: Survey and Experiments. *User Modeling and User-Adapted Interaction* 12, 4, 331–370.

Sammut C., & Webb, G., I. (2017). Encyclopedia of Machine Learning and Data Mining (2nd. ed.). Springer Publishing Company, Incorporated.

Amatriain, X., & Basilico, J. (2016). Past, Present, and Future of Recommender Systems: An Industry Perspective. In *Proceedings of the 10th ACM Conference on Recommender Systems (RecSys' 16)*. Association for Computing Machinery, New York, NY, USA, 211–214.

Gaspar, P., Kompan, M., Koncal, M., & Bielikova, M. (2019). Improving the Personalized Recommendation in the Cold-start Scenarios. In *2019 IEEE International Conference on Data Science and Advanced Analytics* (DSAA), Washington, DC, USA, 606-607.

Walek, B., & Fojtik, V. (2020). A hybrid recommender system for recommending relevant movies using an expert system. In *Expert Systems with Applications.* Vol. 158, 114-152.

Hernández-Rubio, M., Cantador, I. & Bellogín, A. (2019). A comparative analysis of recommender systems based on item aspect opinions extracted from user reviews. In *User Model User-Adapted Interactions*. Springer, No 29, 381–441.

Jain, P.K., Pamula, R., Ansari, S., Sharma, D.K., & Maddala, L. (2019). Airline recommendation prediction using customer generated feedback data. In 2019 4th International Conference on Information Systems and Computer Networks (ISCON), 376-379.

Pradhan, R., Khandelwal, V., Chaturvedi A., & Sharma, D. K. (2020). Recommendation System using Lexicon Based Sentimental Analysis with collaborative filtering. In 2020 International Conference on Power Electronics & IoT Applications in Renewable Energy and its Control (PARC), Mathura, Uttar Pradesh, India, 129-132.

Agarwal, Y., Sharma, D. K., & Katarya, R. (2019). Sentiment/Opinion Review Analysis: Detecting Spams from the good ones! In 2019 4th International Conference on Information Systems and Computer Networks (ISCON), Mathura, India, 557-563.

Garg, S., & Sharma, D.K. (2016). Sentiment Classification of Context Dependent Words. In: Satapathy S., Joshi A., Modi N., Pathak N. (eds). In *Proceedings of International Conference on ICT for Sustainable Development. Advances in Intelligent Systems and Computing*. Springer, Vol 408. 707-715.

Kaššák, O., Kompan, M., & Bieliková, M. (2016). Personalized hybrid recommendation for group of users: Top-N multimedia recommender. Information Processing & Management, 52(3), 459-477.

Shalom, O.S., Berkovsky, S., Ronen, R., Ziklik, E., & Amihood, A. (2015). Data Quality Matters in Recommender Systems. In *Proceedings of the 9th ACM Conference on Recommender Systems (RecSys' 15)*. Association for Computing Machinery, New York, NY, USA, 257–260.

Danilova, V., & Ponomarev, A. (2016). Hybrid Recommender Systems: The Review of State-of-the-Art Research and Applications. In *Proceedings of the 20th Conference of Fruct Association*. 572-578.

Berger, P., & Kompan, M. (2019). User Modelling for Churn Prediction in E-commerce. In IEEE Intelligent System Journal, 34(2), 44-52.

Kotkov, D., Veijalainen, J., & Wang, S. (2018). How does serendipity affect diversity in recommender systems? A serendipity-oriented greedy algorithm. *Computing*.

Su, X., & Khoshgoftaar, T. M. (2009). A survey of collaborative filtering techniques. *Advances in Artificial Intelligence*, 2009:4.

Zheng, Y., Agnani, M., & Singh, M. (2017). Identification of Grey Sheep Users by Histogram Intersection in Recommender Systems. In G. Cong, W.-C. Peng, W. E. Zhang, C. Li, & A. Sun (Eds.), *Advanced Data Mining and Applications*. Springer International Publishing, 148–161.

Aggarwal, C. C. (2016). Advanced Topics in Recommender Systems. In *Recommender Systems: The Textbook*. Springer International Publishing. 411–448.

Felício, C. Z., Paixão, K. V. R., Barcelos, C. A. Z., & Preux, P. (2017). A Multi-Armed Bandit Model Selection for Cold-Start User Recommendation. In *Proceedings of the 25th Conference on User Modeling, Adaptation and Personalization (UMAP' 17)*. Association for Computing Machinery, New York, NY, USA, 32–40.

Teo, Ch. H., Nassif, H., Hill, D., Srinivasan, S., Goodman, M., Mohan, V., & Vishwanathan, S.V.N. (2016). Adaptive, Personalized Diversity for Visual Discovery. In *Proceedings of the 10th ACM Conference on Recommender Systems (RecSys' 16)*. Association for Computing Machinery, New York, NY, USA, 35–38.

Hariri, N., Mobasher, B., & Burke, R. (2015). Adapting to user preference changes in interactive recommendation. In *Proceedings of the 24th International Conference on Artificial Intelligence (IJCAI'15)*. AAAI Press, 4268–4274.

Cunha, T., Soares, C., & de Carvalho, A. C. P. L. F. (2018). CF4CF: recommending collaborative filtering algorithms using collaborative filtering. In *Proceedings of the 12th ACM Conference on Recommender Systems (RecSys' 18)*. Association for Computing Machinery, New York, NY, USA, 357–361.

Ferrari, D. G., & de Castro, L. N. (2012). Clustering Algorithm Recommendation: A Meta-learning Approach. In B. K. Panigrahi, S. Das, P. N. Suganthan, & P. K. Nanda (Eds.), *Swarm, Evolutionary, and Memetic Computing*. Springer Berlin Heidelberg. 143–150.

Edwards, J. A., & Leslie, D. S. (2020). Selecting multiple web adverts: A contextual multi-armed bandit with state uncertainty. *Journal of the Operational Research Society*, 71(1), 100–116.

Tekin, C., & Turgay, E. (2017). Multi-Objective contextual bandits with a dominant objective. 2017 IEEE 27th *International Workshop on Machine Learning for Signal Processing* (MLSP), 1–6.

Koren. Y. (2010). Factor in the neighbors: Scalable and accurate collaborative filtering. ACM Trans. Knowl. Discov. Data 4:1, pp 24.

George T., & Merugu, S. (2005). A scalable collaborative filtering framework based on co-clustering. Fifth IEEE International Conference on Data Mining (ICDM'05), pp. 4.

Lemire, D., & Maclachlan. A. (2007). Slope One Predictors for Online Rating-Based Collaborative Filtering, pp. 5.

Salakhutdinov R., & Mnih, A. (2007). Probabilistic Matrix Factorization. In Proceedings of the 20th International Conference on Neural Information Processing Systems (NIPS'07). Curran Associates Inc., Red Hook, NY, USA, 1257–1264.

Kula, M. (2015). Metadata Embeddings for User and Item Cold-start Recommendations. pp. 8.

Wang, W., & Lu, Y. (2018). Analysis of the Mean Absolute Error (MAE) and the Root Mean Square Error (RMSE) in Assessing Rounding Model. *IOP Conference Series: Materials Science and Engineering*, 324, 12049.

Clarke, Ch. L. A., Kolla, M., Cormack, G. V., Vechtomova, O., Ashkan, A., Büttcher, S., & MacKinnon, I. (2008). Novelty and diversity in information retrieval evaluation. In *Proceedings of the 31st annual international ACM SIGIR conference on Research and development in information retrieval (SIGIR' 08)*. Association for Computing Machinery, New York, NY, USA, 659–666.

Raghavan, V., Bollmann, P., & Jung, G. S. (1989). A critical investigation of recall and precision as measures of retrieval system performance. *ACM Transactions on Information Systems*. 7, 3, 205–229.